\documentclass[12pt,twocolumns,article,nofontune]{IEEEtran}

\usepackage[numbers,sort&compress]{natbib}
\usepackage{url}  
\usepackage{amsthm}
\usepackage{amssymb}
\usepackage{amsfonts}
\usepackage{amsmath}
\usepackage{eso-pic}
\usepackage{graphicx}
\usepackage{tikz}
\usepackage{pgf}
\usepackage{lscape}
\usepackage{fancyvrb}
\usepackage{verbatim}
\usepackage{xfrac}
\usepackage{algpseudocode,algorithm,algorithmicx}
\usepackage{multirow}
\usepackage{proof}
\usepackage{epstopdf}
\usepackage{subfigure}
\usepackage{multicol}
\usetikzlibrary{arrows,automata}
\usetikzlibrary{shapes}

\newcommand{\reffig}[1]{Fig.~\ref{#1}}   

\theoremstyle{plain}

\newtheorem*{corollary*}{Corollary}

\theoremstyle{definition}

\newtheorem*{definition*}{Definition}

\theoremstyle{remark}

\newtheorem*{remark*}{Remark}

\title{Toward Operator-to-Waveform \\5G Radio Access Network Slicing}
\author{\IEEEauthorblockN{
Salvatore D'Oro, Francesco Restuccia, and Tommaso Melodia
\thanks{This paper has been submitted for possible publication in IEEE Communications Magazine. © 2019 IEEE.  Personal use of this material is permitted.  Permission from IEEE must be obtained for all other uses, in any current or future media, including reprinting/republishing this material for advertising or promotional purposes, creating new collective works, for resale or redistribution to servers or lists, or reuse of any copyrighted component of this work in other works.}}\\
\IEEEauthorblockA{Institute for the Wireless Internet of Things, Northeastern University, Boston, USA\vspace{-1cm}}
}

\fontsize{12}{10}\selectfont

\begin{document}
\maketitle
\thispagestyle{plain}
\pagestyle{plain}

\begin{abstract}
Radio access network (RAN) slicing realizes a vision where physical network resources that belong to a specific infrastructure provider can be shared among multiple mobile network operators (MNOs). Existing work in this area has addressed RAN slicing at different levels of network abstractions, but has often neglected the multitude of tightly intertwined inter-level operations involved in real-world slicing systems. For this reason, this article discusses a novel framework for \textit{operator-to-waveform} 5G RAN slicing. In the proposed framework,  slicing operations are treated holistically, including MNO's selection of base stations (BSs) and maximum number of users, down to the waveform-level scheduling of resource blocks. Experimental results show that the proposed framework provides up to 150\% improvement in terms of number of resource blocks that can be used to enable 5G transmission technologies that require coordination and synchronization among BSs.
%
\end{abstract}

\vspace{-0.2cm}
\section{Introduction}

Radio Access Network (RAN) slicing is expected to be a pivotal component of next-generation (5G) networks \cite{rost2017network} and the Internet of Things \cite{restuccia2018securing,restuccia2019big}. RAN slicing leverages virtualization to enable real-time sharing of the physical infrastructure -- owned by an infrastructure provider (IP) -- among  multiple mobile network operators (MNOs). The IP assigns MNOs one or more \textit{slices} of the RAN, each representing a virtual network built on top of the underlying physical RAN. For each slice, the IP specifies the amount of network resources (\textit{e.g.}, base stations (BS), spectrum, transmission power, among others) that can be used by MNOs to provide network services to the Mobile Users (MUs). 

A key aspect of RAN slicing is that MNOs are assigned slices that are strictly independent from one another. In other words, the physical-layer allocation of radio resources (\textit{e.g.}, resource blocks in LTE) is completely up to the MNO, who can allocate them at will based on the proffered demand, pricing schedules, and Quality-of-Experience (QoE) levels. This provides MNOs with a great deal of flexibility, since they can leverage the IP's RAN infrastructure without the need to share business-specific information or low-level scheduling policies with the IP.

Thanks to these core features, RAN slicing has recently attracted considerable interest from academia and industry alike \cite{samdanis2016network,afolabi2018network,bangerter2014networks,GSMA}. This is not at all surprising -- indeed, RAN slicing provides a cost-effective, flexible and efficient solution to core challenges faced by IPs, including (i) the scarcity of networking resources (\textit{e.g.}, spectrum, antennas and BSs); (ii) the need for cost-effective resource allocation strategies; and (iii) the ever-increasing demand for service differentiation through slices tailored to provide services with diverse QoE requirements, \textit{e.g.}, video content delivery, web browsing, real-time surveillance monitoring, among others. Indeed, MNOs can adapt their RAN slice requests to subscribers' requirements and traffic patterns in real-time (\textit{e.g.}, by increasing the resource demand when MUs request high data-rate services, or reducing it in small crowded areas), thus avoiding extra costs due to overbuying of network resources.

By allowing network sharing, IPs can finally overcome the dreaded resource under-utilization issue -- which necessarily comes by using static and exclusive allocation policies and that has plagued previous network generations \cite{underutiliz}. This latter aspect makes RAN slicing a beneficial technology for MNOs -- who are now obliged to compete with each other to provide the best possible service -- and a profitable business model for IPs \cite{GSMA}.  On the other hand, IPs expand their business to the continuously growing market of flexible, high performance and on-demand network deployment for differentiated service typical of 5G networks. 
\begin{figure}[t!]
    \centering
    \includegraphics[width=\columnwidth]{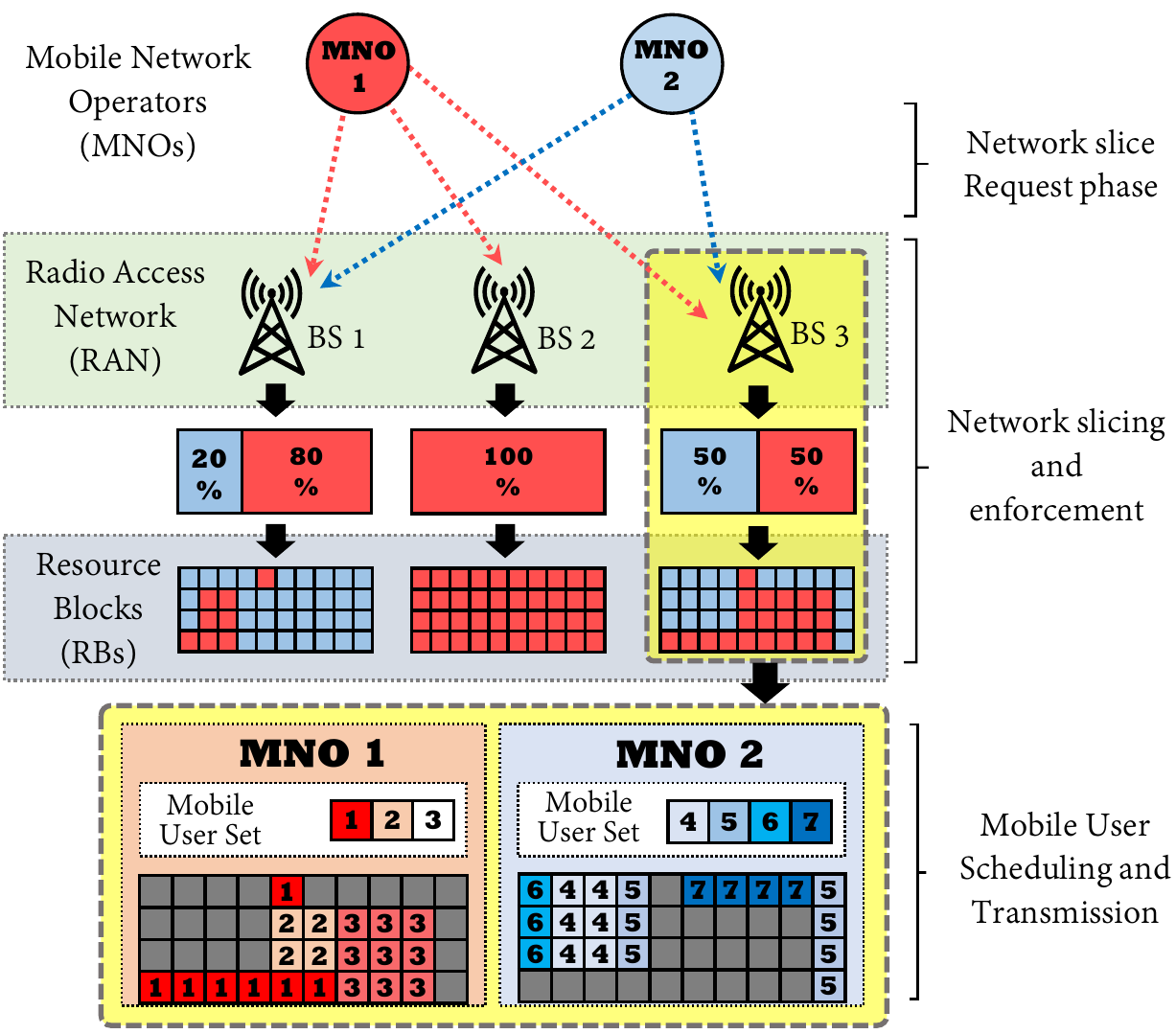}
    \vspace{-0.8cm}
    \caption{An overview of RAN slicing procedures with three BSs and two MNOs. We focus on BS $3$ whose resources are sliced among MNO $1$ and $2$ serving three and four MUs, respectively.}
    \label{fig:scheme}
 \end{figure}
 
Despite being one of the most promising 5G technologies, RAN slicing and its application to next-generation networks do not come without key challenges \cite{afolabi2018network}. To be considered an integral part of future 5G networks, RAN slicing will have to be simultaneously \textit{cost-effective}, \textit{easy to use}, as well as providing \textit{high-level network performances}. Poor performance, high prices, and a cumbersome slicing interface may indeed discourage potential MNOs from joining the RAN slicing market. Similarly, high maintenance and management costs may make RAN slicing an unprofitable business for IPs. Furthermore, to obtain the tens of Gbps \cite{parvez2018survey}-level data rates envisioned for 5G, deployed RAN slices will have to support advanced transmission technologies such as Coordinated Multi-Point (CoMP), multi-user and massive MIMO \cite{bangerter2014networks}. These technologies, however, have strict timing requirements (in the order of few milliseconds), require in-depth understanding of low-level details of the physical network and demand for coordination among BSs, which results in increased networking and operational costs.

From the above discussion, it is clear that the ability of IPs to efficiently trade-off between business and networking aspects is a cornerstone to the success of RAN slicing. On the business side, this crucial trade-off can only be reached  when IPs will implement \textit{holistic, operator-to-waveform} 5G slicing solutions,  where MNOs are enabled to express high-level networking need, yet obtain and retain full control of the right amount of physical and spectrum resources needed to achieve it. Simply put, we envision a framework where an MNO can ask the IP something in the tune of ``There is an event between 7:30 and 10:30 at TD Garden in Boston, where approximately 1000 spectators will be present. I expect each user to generate traffic of 10 MB/s." The framework will then \textit{automatically} generate the physical resources to achieve this goal, and leave to the MNO the task to handle the waveform-level scheduling operations according to his QoE business plan, \textit{e.g.}, ``premium'' users get maximum 20MB/s data rate, whereas ``basic'' users get instead only 5MB/s.   

Although existing work has already tackled slicing-related problems, to the best of our knowledge a unified operator-to-waveform 5G RAN slicing framework is still missing to this day. For this reason, in this work we discuss the road ahead to achieve this ambitious yet crucial goal. Specifically, we identify the architecture design features and we discuss them in detail. Our focus is also toward the identification of crucial aspects that are of extreme importance to enable advanced communication technologies envisioned and strongly utilized in 5G networks.
\begin{figure}[t!]
    \centering
    \includegraphics[width=\columnwidth]{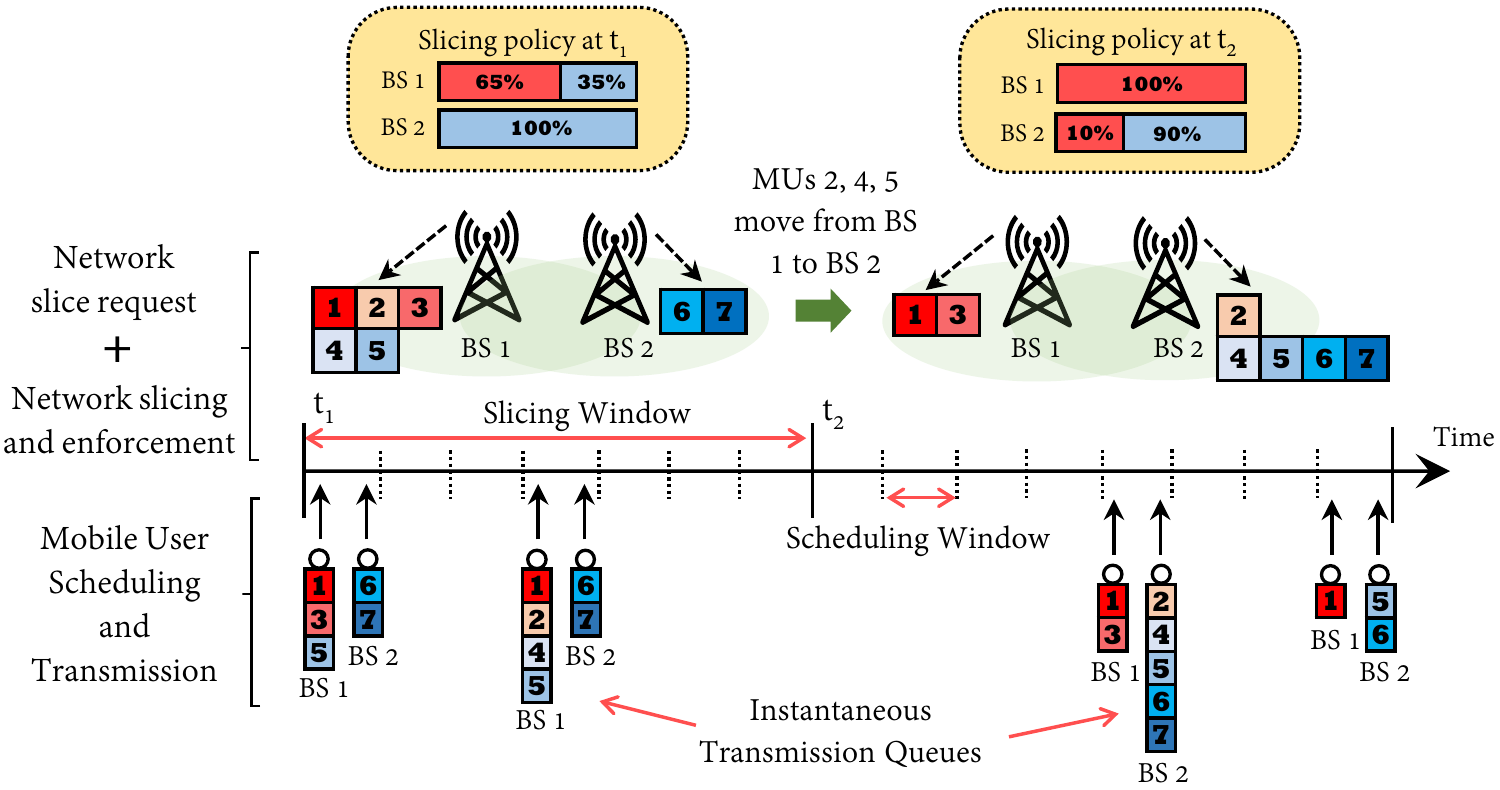}
    \vspace{-0.8cm}
    \caption{Timing aspects in RAN slicing and scheduling with two BSs and seven MUs. It is shown that MNOs adapt their slicing strategies to MUs mobility patterns (long time-scale), while transmission queues rapidly vary in time (short time-scale).}
    \label{fig:time}
\end{figure}

The remainder of this article is organized as follows: Section \ref{sec:glance} gives an overview of the RAN slicing problem, with particular emphasis on its application to 5G networks. Section \ref{sec:architecture} discusses the architecture design of a unified RAN slicing ecosystem spanning across MNOs and IPs domains. The effectiveness of the proposed architecture is assessed in Section \ref{sec:experiments}, and concluding remarks are given in Section \ref{sec:conclusions}.

\vspace{-0.4cm}
\section{RAN Slicing: \\ Requirements and Challenges} \label{sec:glance}

\reffig{fig:scheme} provides an overview of the main procedures involved in RAN slicing. In a nutshell, we can divide the core operations into three broad yet logically distinct phases. First, each MNO  declares the desired slice configuration (\textit{slice request generation} phase), \textit{e.g.}, how many resources are required, which BSs should be included in the slice, the desired QoE level, among others. Then, the IP  decides which MNO requests can be accommodated and how many networking resources should be allocated to each slice (\textit{network slice generation} phase). Virtual RAN slices are generated by allocating physical networking resources (\textit{e.g.}, resource blocks) to each MNO. Finally, each MNO takes control over the slice to schedule MUs' transmissions (\textit{MUs scheduling and transmission}  phase). \vspace{-0.2cm}

\vspace{-0.25cm}
\subsection*{The need for a unifying slicing system} 

At first blush, RAN slicing might look a minor variation of already well-established cloud-related slicing technologies \cite{ordonez2017network,blenk2016survey}. However, RAN slicing is an intrinsically different problem, since (i) spectrum is a scarce resource for which over-provisioning is not possible; (ii) the network capacity is dynamic and heavily depends on rapidly fluctuating RAN-specific factors such as location of both MUs and BSs; (iii) electromagnetic interference rapidly varies over time and threatens orthogonality across different slices belonging to different MNOs; and (iv) the agreements with MNOs usually impose stringent requirements on the subscriber QoE, which however strongly depend on channel conditions and MU mobility patterns. 

In other words, we need radically novel approaches where heterogeneous yet interdependent tasks such as resource virtualization (high-level), interference management and data transmission (low-level) are jointly considered to develop a unified slicing framework. \vspace{-0.2cm}

\vspace{-0.25cm}
\subsection*{Different time-scales}

A crucial aspect is that the three phases in  \reffig{fig:scheme} work at different temporal scales. Indeed, as shown in \reffig{fig:time}, while the request and generation of virtual RAN slices take place within a \textit{slicing window} whose overall duration fluctuates from few tens of milliseconds to several months \cite{marquez2018should}, MU scheduling and data transmission follow more minute dynamics. For example, scheduling and data transmission in 4/5G networks is performed on time slots lasting one millisecond only -- the duration of a Transmission Time Interval (TTI). This is because scheduling requires network state information which is available only at network runtime (\textit{e.g.}, MU position, channel state information (CSI), size of scheduling queues, and so forth). 

\vspace{-0.2cm}
\subsection*{Enabling emerging multi-BSs wireless technologies}

The promise of high-speed 5G communications heavily relies on key multi-BS technologies including CoMP, Massive MIMO, and beamforming \cite{bangerter2014networks}, which ground their effectiveness on the tight cooperation and coordination among the different serving BSs. Yet, surprisingly, traditional slicing algorithms neglect to consider this core aspect, leading to network performance degradation. Only recently \cite{DoroINFOCOM} it became evident that designing fine-grained RAN slicing algorithms that partition network resources to create a flourishing ground for cooperation- and coordination-based communications will become more and more crucial in the years to come.

\begin{figure}[h!]
    \centering
    \includegraphics[width=0.95\columnwidth]{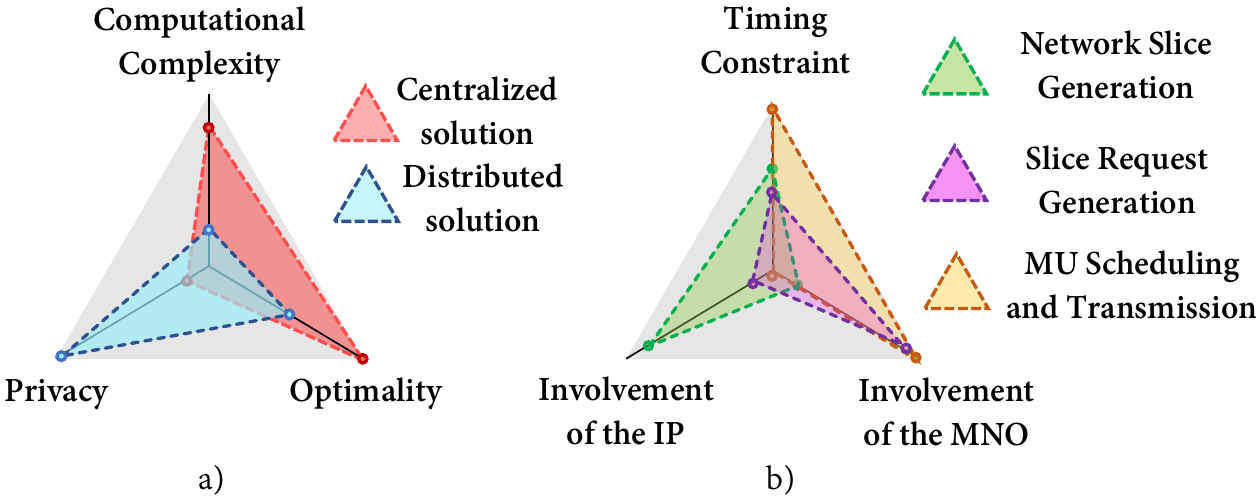}
    \vspace{-0.4cm}
    \caption{Triple constraint diagrams summarizing the RAN slicing optimization constraints. On the left, we show (a) the complexity vs. optimality vs. privacy trade-offs, while on the right  we show (b) features and requirements for each phase of the RAN slicing problem.\vspace{-0.5cm}}
    \label{fig:triple}
 \end{figure}

\vspace{-0.2cm}
\subsection*{Complexity vs. Optimality vs. Privacy}

Another main challenge in designing an operator-to-waveform RAN slicing framework is the need to find an efficient balance among \textit{complexity}, \textit{optimality}, and \textit{privacy}. This balance can be represented with the triple constraint in the left-hand side \reffig{fig:triple}.

Figure \ref{fig:triple} shows that centralized approaches generally produce optimal slicing policies to the detriment of privacy. Indeed, these strategies assume that the IP is aware of MNOs' relevant information, \textit{e.g.}, monetary budget, subscribers' number and position, business strategies, among others. MNOs, however, are extremely reluctant to disclose such sensitive information in real-world scenarios. Also, it has been shown that centralized formulations of the RAN slicing problem are provably NP-hard \cite{DoroINFOCOM}. For this reason, \textit{distributed approaches} are more desirable \cite{DoroTNET}. 
 
Addressing these triple constraint requires deep understanding of the dynamics governing each phase of the RAN slicing problem. As shown in \reffig{fig:triple}, the \textit{slicing request generation} phase has mild timing constraints and mostly involves MNOs only. The \textit{network slicing generation} phase, instead, requires direct control of the IP to efficiently apportion the available network resource while reducing intra-MNO interference and enabling coordination-based communications. Lastly, the \textit{MU scheduling and transmission} phase is individually controlled by each MNO, does not involve the IP at all yet has very strict timing constraints. 

The above discussion suggests that \textit{the RAN slicing problem has diverse and generally opposing requirements}. Recent research work has suggested to let MNOs handle the \textit{slicing request generation} and \textit{MU scheduling and transmission} phases, while leaving the \textit{network slicing generation} phase to the IP \cite{DoroTNET}. Conversely, the \textit{network slicing generation} phase is left to the IP which possess the global network view required to implement fine-grained RB allocation \cite{DoroINFOCOM}. This approach offers a balanced trade-off between centralization, complexity, optimality, and privacy, yet also captures requirements and distinctive interactions among IPs and MNOs.

\vspace{-0.25cm}
\section{System Architecture} \label{sec:architecture}

\begin{figure*}[t]
    \centering
    \includegraphics[width=\textwidth]{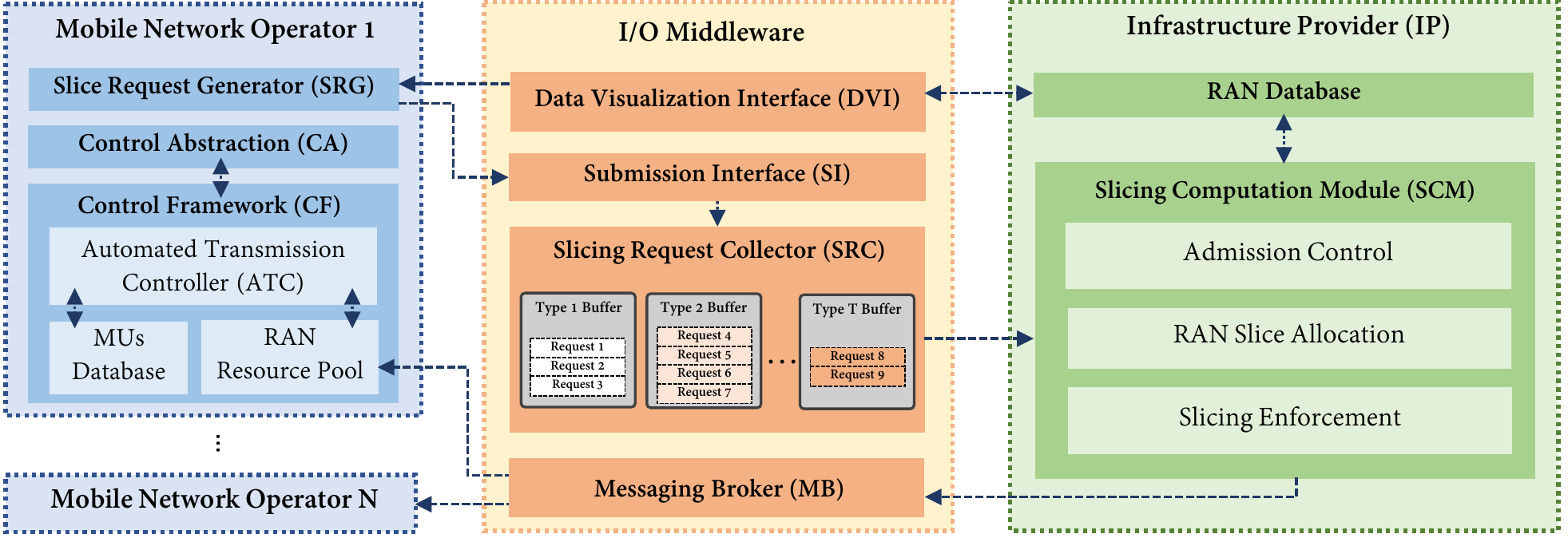}
    \vspace{-0.8cm}
    \caption{The architecture of the proposed three-tier RAN slicing framework.}
    \label{fig:architecture}
 \end{figure*}

The proposed operator-to-waveform RAN slicing framework is illustrated in \reffig{fig:architecture}. It consists of a three-tier architecture. The \textit{IP space} enables IPs to (i) process MNOs' requests, and (ii) generate RAN slices according to such requests. The \textit{MNO space} concerns functionalities required by MNOs to (i) visualize the available network resources, (ii) generate slice requests, and (iii) control the obtained RAN slices and network resources. Finally, the \textit{I/O Middleware} enables and regulates interactions among MNOs and between every MNO and the IP.

\vspace{-0.5cm}
\subsection{The I/O Middleware Tier} \label{sec:io}

\textbf{Database Visualization Interface (DVI):} the DVI provides MNOs access to the list of available BSs, their location, leasing price and coverage details, as well as useful metrics such as provided QoS (\textit{e.g.}, average QoS Class Identifier (QCI) and throughput) and the current congestion and utilization levels of each BS, among others. 

\textbf{Submission Interface (SI):} the slicing requests generated by the \textit{Slice Request Generator} block (Section \ref{sec:MNO}) are submitted to the IP through the SI. The design of the SI can rely upon web-based services \cite{DoroTNET} or client-based software with dedicated GUIs to facilitate the submission process \cite{DoroINFOCOM}.
    
\textbf{Slicing Request Collector (SRC):} this block collects MNOs' requests submitted through the SI. As discussed earlier, MNOs are expected to submit RAN slice requests that substantially differ one from another. For example, some MNOs might be willing to include a specific BS to their slice \cite{DoroTNET, DoroINFOCOM}. Instead, other MNOs could be interested in covering a specific point-of-interest (PoI) (\textit{e.g.}, landmarks, schools, theaters, and other locations that might be of interest for MNOs) disregarding which BS is actually used to serve subscribers \cite{afolabi2018network}. Requests are organized into multiple classes, where each class contains homogeneous requests in terms of requested RAN configuration and QoS. Each class is then assigned to a dedicated buffer where all requests for that specific class are stored.

\textbf{Messaging Broker (MB):} this block is used by the IP to send slice admission/rejection notifications to each requesting MNO and, if accepted, to specify which BSs have been included in the current slice, the amount of networking resources that have been assigned to the MNO, and the price to be paid to rent the physical infrastructure. When a slice request is accepted, this information is stored inside the \textit{Resource Pool} (RP) (Section \ref{sec:MNO}) residing within the MNO space.
Together with the DVI and the RP, this block is used by MNOs as feedback channel to gather information on the current slicing configuration. Section \ref{sec:MNO} will show that such feedback is crucial to realize the decentralized vision where MNOs autonomously compute their slicing requests. 

\vspace{-0.41cm}
\subsection{The IP Tier} \label{sec:no}

\textbf{RAN Database:} it contains detailed information on the network topology (\textit{e.g.}, position of each BS and its interfering BSs), leasing price and the amount of available resources at each BS (\textit{e.g.}, number of antennas, operational frequencies and RBs availability). Furthermore, the database has information on coverage and performance properties of each BS (\textit{e.g.}, whether or not a BS covers a given PoI and what are the congestion and QCI levels at specific points of the network). 
    
The RAN database contains sensitive information that the IP uses for maintenance and monitoring purposes (\textit{e.g.}, how many resources are allocated to each MNO, MNO preferences toward a specific subset of BS and historical data on the RAN slice requests received in the past). Since the IP might be reluctant to share such abundant information with MNOs, they can visualize information contained in the RAN database through the DVI only. This way, the DVI obfuscates any IP-related sensitive information.

\textbf{Slicing Computation Module (SCM):} this block is responsible for computing optimal RAN slicing strategies based on MNOs' slice requests received by the \textit{Slicing Request Collector} (SRC) module.
This block also takes care of (i) deciding which requests can be admitted; (ii) computing a slicing policy to allocate the available resources to the admitted MNOs; and (iii) enforce the computed slicing policy on the underlying physical RAN.  The IP specifies a desired objective function (\textit{e.g.}, throughput maximization, congestion \cite{DoroTNET} and interference \cite{DoroINFOCOM} minimization) subject to one or more constraints (\textit{e.g.}, guarantee Service Level Agreements (SLAs) or minimum QoS level). The SCM computes a slicing strategy that meets IP's directives via the following three procedures:

$\bullet$ \textit{Admission Control (AC):} this procedure determines which requests can be admitted and which are to be rejected. If a MNO submits an unfeasible request (\textit{e.g.}, it demands an excessive minimum data-rate guarantee for subscribers in a very congested area, or a large amount of resources on a BS already assigned to other MNOs), the AC procedure refuses the request and notifies the corresponding MNO through the MB;

$\bullet$ \textit{RAN Slice Allocation (RSA):} admitted requests are assigned with a portion of the available networking resources commensurate with the amount of resources they requested. An example is illustrated in \reffig{fig:scheme} where MNO 1 obtains $20\%$, $100\%$ and $50\%$ of the available resources on BS $1$, $2$ and $3$, respectively;

$\bullet$ \textit{Slicing Enforcement (SE):} this procedure completes the RAN slicing process. An example is depicted in \reffig{fig:scheme} where we show how the slicing policy $(50\%,50\%)$ generated by the RSA procedure for BS $3$ is then enforced over the RB grid by assigning specific RBs to the two requesting MNOs. This procedure is also aware of the network topology, with particular focus on adjacency (or interference) matrices \cite{DoroINFOCOM}. These matrices provide extremely useful information with respect to adjacency among close BSs, thus enabling algorithms for technologies such as MIMO, beamforming and CoMP transmissions. Hence, these matrices are used to allocate the same RBs (in the time/frequency domain) to the same MNO when BSs are close enough to interfere among themselves \cite{DoroINFOCOM}.

We remark that the above three procedures are tightly intertwined -- and their impact on network performance and IP's profit is extremely significant. As such, these procedures must be tailored to admit as many slice requests as possible, improve network performance (\textit{e.g.}, maximize throughput and spectral efficiency, minimize interference) and enable advanced transmission technologies.

\vspace{-0.5cm}
\subsection{The MNO Tier} \label{sec:MNO}

\textbf{Slice Request Generator (SRG):} after having visualized information related to the underlying RAN  via the DVI (Section \ref{sec:no}), MNOs formulate their request through the SRG -- by specifying, for example, which BSs should be included in the slice, which PoIs should be covered, the required number of resources and minimum QoS levels, among others. The generated requests are then submitted through the \textit{Submission Interface} (Section \ref{sec:no}). As remarked earlier, it is important that this process is autonomously executed by each MNO, so that they can adapt their RAN slice to network changes and lower the computational burden on the IP. Moreover, by granting MNOs access to cumulative metrics such as overall network congestion, interference measurements and total amount of resources allocated to the RAN slice, it is possible to achieve distribute RAN slicing \cite{DoroTNET}. This eliminates the need for any coordination mechanism among MNOs and preserves their privacy. In view of the above, the SRG gathers such cumulative information from the DVI and RAN Resource Pool and formulates a new RAN slice request accordingly. Lastly, the generated request is submitted to the IP through the \textit{Submission Interface (SI)}.

\textbf{Control Abstraction (CA):} the CA allows the MNO to define high-level control directives to optimize specific objective functions (\textit{e.g.}, transmission rate, latency, spectral efficiency) subject to one or more constraints (\textit{e.g.}, minimum data-rate, maximum end-to-end delay and transmission power). This block operates as an abstraction layer hiding low-level network details -- such as resources allocated to the RAN slice and MUs' position, number and generated traffic, among others -- to the MNO. 

\textbf{Control Framework (CF):} this block transforms MNO's control directives defined in the CA (\textit{e.g.}, maximize spectral efficiency while guaranteeing a minimum data-rate to each MU) into low-level resource allocation policies to serve MUs. The CF consists of the three following elements:

$\bullet$ \textit{MU Database:} this database -- akin to the Home Subscriber Server (HSS) already used in current 4G cellular networks -- contains information on MNO's subscribers such as position, QoS requirements, identities and addressing. MNOs use it to (i) determine which BSs should be included in their slice; and (ii) how many resources are needed to serve MUs and satisfy their QoS requirements;

$\bullet$ \textit{RAN Resource Pool:} as soon as the MNO is granted with one or more RAN slices, this block collects information on networking resources assigned to each slice (\textit{e.g.}, operational frequencies, RBs, antennas, and transmission power, among others). This way, MNOs are kept aware of which resources can be utilized to provide network services to mobile subscribers;

$\bullet$ \textit{Automated Transmission Controller (ATC):} this block provides the MNO with an abstraction layer that hides all low-level network details (\textit{e.g.}, MU position and CQI, employed modulation, resources allocated to the RAN slice and power budget at each BS). This is achieved by converting high-level directives into optimization algorithms that optimally allocate RAN slice resources and meet MNO's requirements and directives \cite{guan2018wnos}. It follows that any MNO may efficiently and automatically control RAN slices without any in-depth knowledge of the underlying physical network as well as resource allocation and optimization algorithms.

\vspace{-0.2cm}
\section{Multi-BS Coordination: A Case Study} \label{sec:experiments}

\begin{figure}[t!]
    \centering
    \includegraphics[width=0.95\columnwidth]{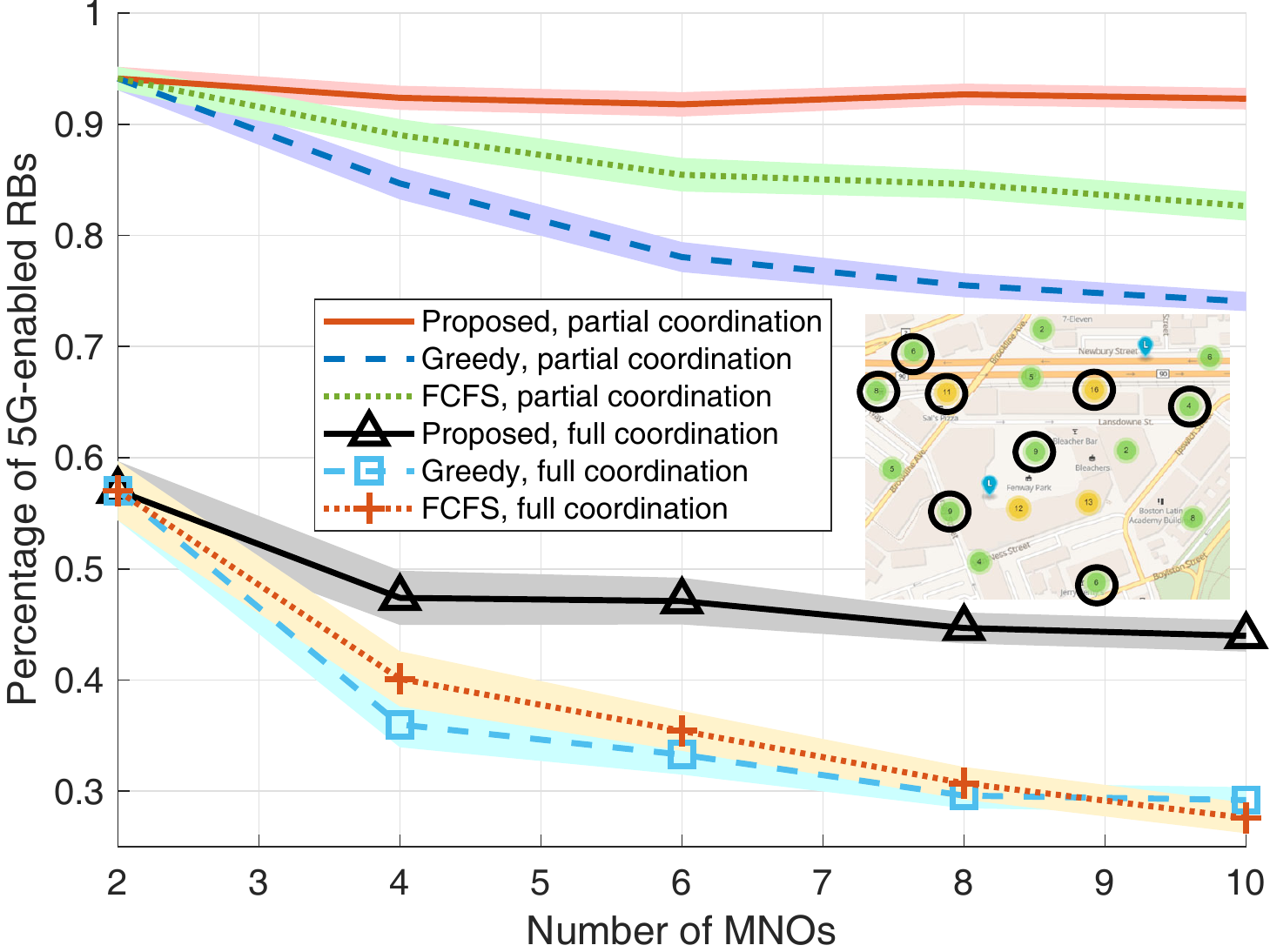}
    \vspace{-0.4cm}
    \caption{Physical RAN deployment (selected BSs are highligted with black circles) and percentage of RBs that enable coordinated multi-BS wireless technologies under different RAN slicing strategies. Curves show average values and shaded areas display $95\%$ confidence intervals over $200$ independent simulation runs.}
    \label{fig:plot}
\end{figure}

We consider a cluster of $8$ BSs deployed in Boston (USA) extracted from the OpenCellID database (https://opencellid.org), which is shown in \reffig{fig:plot}. We consider an LTE deployment where RAN slicing is performed at the RB level \cite{DoroINFOCOM}. Each BS has $50$ RBs, and MNOs distributively submit RAN slice requests through the SI (Section \ref{sec:io}), aiming at minimizing the network congestion and cost per slice \cite{DoroTNET}. Requests are collected by the SRC and converted into effective slicing strategies by the SCM (Section \ref{sec:no}) which slices network resources so that interference is minimized and multi-BS coordination is maximized \cite{DoroINFOCOM}. 

In \reffig{fig:plot} we show how RAN slicing policies affect 5G-related transmission technologies such as MIMO and CoMP. Specifically, \reffig{fig:plot} shows the percentage of RBs that can be leveraged for multi-BS wireless transmissions for different slicing enforcement strategies, \textit{i.e.}, \textit{shared RBs}. We compare the approach proposed in \cite{DoroINFOCOM} -- which maximizes the chances that a RB can be used to serve MUs via two (or more) BSs in close proximity -- with greedy (\textit{i.e.}, RBs are first allocated to those MNOs that have been assigned with the the largest amount of RBs) and First Come First Serve (FCFS) (\textit{i.e.}, RBs are sequentially allocated to MNOs according to the temporal order of their RAN slice request submission).
We notice that the number of shared RBs decreases with the number of MNOs for both \textit{full coordination} (\textit{i.e.}, all BSs can serve the same MU in the same RB) and \textit{partial coordination} (\textit{i.e.}, at least two BSs can serve the same MU in the same RB). The approach in \cite{DoroINFOCOM} provides a performance gain of up to $150\%$ in terms of number of RBs that can be used to enable 5G technologies that require coordination.

\vspace{-0.55cm}
\section{Conclusions} \label{sec:conclusions}

We discussed a unified framework for operator-to-waveform 5G RAN slicing, which allows control from the MNO's selection of base stations and maximum number of users to the waveform-level scheduling of resource blocks. Experimental results show that our framework allows for up to 150\% improvement in terms of number of RBs that can be used to enable coordination-based 5G technologies, such  as MIMO  and  CoMP. 

\vspace{-0.35cm}
\footnotesize
\bibliographystyle{IEEEtran}
\bibliography{bibl} 

\begin{thebibliography}{10}
\providecommand{\url}[1]{#1}
\csname url@samestyle\endcsname
\providecommand{\newblock}{\relax}
\providecommand{\bibinfo}[2]{#2}
\providecommand{\BIBentrySTDinterwordspacing}{\spaceskip=0pt\relax}
\providecommand{\BIBentryALTinterwordstretchfactor}{4}
\providecommand{\BIBentryALTinterwordspacing}{\spaceskip=\fontdimen2\font plus
\BIBentryALTinterwordstretchfactor\fontdimen3\font minus
  \fontdimen4\font\relax}
\providecommand{\BIBforeignlanguage}[2]{{%
\expandafter\ifx\csname l@#1\endcsname\relax
\typeout{** WARNING: IEEEtran.bst: No hyphenation pattern has been}%
\typeout{** loaded for the language `#1'. Using the pattern for}%
\typeout{** the default language instead.}%
\else
\language=\csname l@#1\endcsname
\fi
#2}}
\providecommand{\BIBdecl}{\relax}
\BIBdecl

\bibitem{rost2017network}
P.~Rost, C.~Mannweiler, D.~S. Michalopoulos, C.~Sartori, V.~Sciancalepore,
  N.~Sastry, O.~Holland, S.~Tayade, B.~Han, D.~Bega \emph{et~al.}, ``{Network
  Slicing to Enable Scalability and Flexibility in 5G Mobile Networks},''
  \emph{IEEE Communications Magazine}, vol.~55, no.~5, pp. 72--79, 2017.

\bibitem{restuccia2018securing}
F.~Restuccia, S.~D'Oro, and T.~Melodia, ``{Securing the Internet of Things in
  the Age of Machine Learning and Software-defined Networking},'' \emph{IEEE
  Internet of Things Journal}, vol.~5, no.~6, pp. 4829--4842, 2018.

\bibitem{restuccia2019big}
F.~Restuccia and T.~Melodia, ``{Big Data Goes Small: Real-Time Spectrum-Driven
  Embedded Wireless Networking Through Deep Learning in the RF Loop},''
  \emph{Proc. of IEEE Conference on Computer Communications (INFOCOM)}, 2019.

\bibitem{samdanis2016network}
K.~Samdanis, X.~Costa-Perez, and V.~Sciancalepore, ``{From Network Sharing to
  Multi-tenancy: The 5G Network Slice Broker},'' \emph{IEEE Communications
  Magazine}, vol.~54, no.~7, pp. 32--39, 2016.

\bibitem{afolabi2018network}
I.~Afolabi, T.~Taleb, K.~Samdanis, A.~Ksentini, and H.~Flinck, ``Network
  slicing and softwarization: A survey on principles, enabling technologies,
  and solutions,'' \emph{IEEE Communications Surveys \& Tutorials}, vol.~20,
  no.~3, pp. 2429--2453, 2018.

\bibitem{bangerter2014networks}
B.~Bangerter, S.~Talwar, R.~Arefi, and K.~Stewart, ``{Networks and Devices for
  the 5G Era},'' \emph{IEEE Communications Magazine}, vol.~52, no.~2, pp.
  90--96, 2014.

\bibitem{GSMA}
\BIBentryALTinterwordspacing
{GSMA Association}, ``Network slicing: Use case requirements,'' 2018. [Online].
  Available: \url{https://tinyurl.com/y3bb3x4h}
\BIBentrySTDinterwordspacing

\bibitem{underutiliz}
X.~{Zhou} and L.~{Chen}, ``{Demand Shaping in Cellular Networks},'' \emph{IEEE
  Transactions on Control of Network Systems}, vol.~6, no.~1, pp. 363--374,
  March 2019.

\bibitem{parvez2018survey}
I.~Parvez, A.~Rahmati, I.~Guvenc, A.~I. Sarwat, and H.~Dai, ``{A Survey on Low
  Latency Towards 5G: RAN, Core Network and Caching Solutions},'' \emph{IEEE
  Communications Surveys \& Tutorials}, vol.~20, no.~4, pp. 3098--3130, 2018.

\bibitem{ordonez2017network}
J.~Ordonez-Lucena, P.~Ameigeiras, D.~Lopez, J.~J. Ramos-Munoz, J.~Lorca, and
  J.~Folgueira, ``{Network Slicing for 5G with SDN/NFV: Concepts,
  Architectures, and Challenges},'' \emph{IEEE Communications Magazine},
  vol.~55, no.~5, pp. 80--87, 2017.

\bibitem{blenk2016survey}
A.~Blenk, A.~Basta, M.~Reisslein, and W.~Kellerer, ``Survey on network
  virtualization hypervisors for software defined networking,'' \emph{IEEE
  Communications Surveys \& Tutorials}, vol.~18, no.~1, pp. 655--685, 2016.

\bibitem{marquez2018should}
C.~Marquez, M.~Gramaglia, M.~Fiore, A.~Banchs, and X.~Costa-Perez, ``{How
  Should I Slice My Network?: A Multi-Service Empirical Evaluation of Resource
  Sharing Efficiency},'' in \emph{Proceedings of the 24th Annual International
  Conference on Mobile Computing and Networking}.\hskip 1em plus 0.5em minus
  0.4em\relax ACM, 2018, pp. 191--206.

\bibitem{DoroINFOCOM}
S.~D'Oro, F.~Restuccia, A.~Talamonti, and T.~Melodia, ``{The Slice Is Served:
  Enforcing Radio Access Network Slicing in Virtualized 5G Systems},'' in
  \emph{Proc. of IEEE International Conference on Computer Communications
  (INFOCOM)}, April 2019.

\bibitem{DoroTNET}
S.~D'Oro, F.~Restuccia, T.~Melodia, and S.~Palazzo, ``{Low-Complexity
  Distributed Radio Access Network Slicing: Algorithms and Experimental
  Results},'' \emph{IEEE/ACM Transactions on Networking}, vol.~26, no.~6, pp.
  2815--2828, 2018.

\bibitem{guan2018wnos}
Z.~Guan, L.~Bertizzolo, E.~Demirors, and T.~Melodia, ``{WNOS: An
  optimization-based wireless network operating system},'' in \emph{Proceedings
  of the Eighteenth ACM International Symposium on Mobile Ad Hoc Networking and
  Computing}.\hskip 1em plus 0.5em minus 0.4em\relax ACM, 2018, pp. 241--250.

\end{thebibliography}

\vspace*{-2\baselineskip}
\begin{IEEEbiographynophoto}
    {Salvatore D'Oro} (M'16) is a research associate scientist at the Institute for the Wireless Internet of Things, Northeastern University (Boston, USA). He received the PhD degree from the University of Catania in 2015. His research interests include game-theory, optimization, learning and their applications to telecommunication networks.
    
    \vspace{0.1cm}
    \noindent
  \textbf{Francesco Restuccia} (M'16) is an Associate Research Scientist at the Institute for the Wireless Internet of Things, Northeastern University (Boston, USA). Dr. Restuccia's research interests lie in the  modeling, analysis, and experimental evaluation of wireless networked systems, with applications to pervasive and mobile computing and the Internet of Things.
  
  \vspace{0.1cm}
  \noindent
 \textbf{Tommaso Melodia} (F'18) received the Ph.D. degree in Electrical and Computer Engineering from the Georgia Institute of Technology in 2007. He is the William Lincoln Smith Chair Professor at Northeastern University. He is the Director of Research for the PAWR Project Office. Prof. Melodia's research focuses on modeling, optimization, and experimental evaluation of wireless networked systems. Prof. Melodia is an Associate Editor for the IEEE Transactions on Wireless Communications, the IEEE Transactions on Mobile Computing, the IEEE Transactions on Biological, Molecular, and Multi-Scale Communications.
\end{IEEEbiographynophoto}

\end{document}